\documentstyle[12pt]{article}
\def\np{Nucl. Phys.}
\def\pl{Phys. Lett.}
\def\pr{Phys. Rev.}
\def\prl{Phys. Rev. Lett.}
\def\cmp{Comm. Math. Phys.}
\def\calF{{\cal F}}

\begin{document}
\baselineskip = 20pt

\def\footnotefont{\tenpoint}
\def\figures{\centerline{{\bf Figure
Captions}}\medskip\parindent=40pt%
\def\fig##1##2{\medskip\item{FIG.~##1.  }##2}}
\newwrite\ffile\global\newcount\figno \global\figno=1
\def\fig{fig.~\the\figno\nfig}
\def\nfig#1{\xdef#1{fig.~\the\figno}%
\writedef{#1\leftbracket fig.\noexpand~\the\figno}%
\ifnum\figno=1\immediate\openout\ffile=figs.tmp\fi\chardef\wfile=
\ffile%
\immediate\write\ffile{\noexpand\medskip\noexpand\item{Fig.\
\the\figno. }
\reflabeL{#1\hskip.55in}\pctsign}\global\advance\figno by1\findarg}

\parindent 25pt
\overfullrule=0pt
\tolerance=10000
\def\Re{\rm Re}
\def\Im{\rm Im}
\def\titlestyle#1{\par\begingroup \interlinepenalty=9999
     \fourteenpoint
   \noindent #1\par\endgroup }
\def\tr{{\rm tr}}
\def\Tr{{\rm Tr}}
\def\half{{\textstyle {1 \over 2}}}
\def\calt{{\cal T}}
\def\ie{{\it i.e.}}
\def\np{Nucl. Phys.}
\def\pl{Phys. Lett.}
\def\pr{Phys. Rev.}
\def\prl{Phys. Rev. Lett.}
\def\cmp{Comm. Math. Phys.}
\def\quart{{\textstyle {1 \over 4}}}
\def\RR{${\rm R}\otimes{\rm R}~$}
\def\NSNS{${\rm NS}\otimes{\rm NS}~$}
\def\calf{${\cal F}$}

\baselineskip=14pt
\pagestyle{empty}
{\hfill DAMTP/96-34}
\vskip 0.1cm
{\hfill hep-th/9604091}
\vskip 0.4cm
\centerline{ LIGHT-CONE SUPERSYMMETRY AND D-BRANES.}
\vskip 1cm
 \centerline{ Michael B.  Green\footnote{M.B.Green@damtp.cam.ac.uk}
 and Michael Gutperle\footnote{M.Gutperle@damtp.cam.ac.uk}}
\vskip 0.3cm
\centerline{DAMTP, Silver Street,}
\centerline{ Cambridge CB3 9EW, UK.}
\vskip 1.4cm
\centerline{ABSTRACT}
\vskip 0.3cm
$D$-brane boundary states for type II superstrings are constructed
by enforcing
the conditions that preserve half of the  space-time supersymmetry.   A
light-cone coordinate frame is used where time is identified as one
of the
coordinates transverse to the  brane's (euclidean)  world-volume so
that the
$p$-brane is treated as a $(p+1)$-instanton.    The boundary states
have the
superspace  interpretation of top or bottom states in a light-cone string
superfield.  The presence of a non-trivial open-string boundary
condensate  give rise to  the familiar $D$-brane source terms that
determine the (linearized) Born--Infeld-like effective actions for
$p$-branes and
the (linearized) equations of motion for the massless fields
implied by the
usual $p$-brane ansatze.
The `energy' due to closed string exchange
between separate
$D$-branes  is calculated  (to lowest order in the string coupling) in
situations with pairs of parallel, intersecting as well as
orthogonal branes --  in
which case the unbroken supersymmetry may be  reduced.
Configurations of more than two branes  are also
considered in situations in which the supersymmetry is reduced to
$1/8$ or $1/16$ of the full amount.
The Ward identities resulting from  the non-linearly realized broken space-time
supersymmetry in the presence of a $D$-brane are also discussed.

 \vfill\eject
\pagestyle{plain}
\setcounter{page}{1}

\section{Introduction}

The description of  the $p$-branes of the \RR sector of type II
superstring
theories in terms of $D$-branes
\cite{polchina,polchinskinotes,wittenb} has provided a
string-theoretic
interpretation  of the world-volume theories of these objects.
 According to  the $D$-brane description the \RR sector $p$-branes  are
described by configurations of open superstrings with end-points
tethered on a
$(p+1)$-dimensional hypersurface embedded in ten dimensions.   The
coordinates of the open-string end-points are
restricted to the plane $X^i = y^i$, where the directions labelled
$i$ are
transverse to the brane world-volume ($i =p+1, \cdots, 9$).  The
dynamics of
the brane is  therefore prescribed by the open superstring theory
with Neumann
boundary conditions in the directions labelled by $\alpha =0,\dots,
p$ and
Dirichlet boundary conditions in the transverse directions.
The usual kind of  \lq effective' world-volume field theory emerges
 as an
approximation to this \lq underlying'  open superstring theory.  The
world-volume actions  that describe the various  branes
\cite{hughespolchinskia,duffa} are
generally non-renormalizable $(p+1)$-dimensional  supersymmetric
field theories
that are generalizations of the Nambu--Goto action \cite{leigha} and the
covariant  superstring action  to higher dimensions.
    The
fact that  such $D$-branes are described by {\it open} superstring
theory means
that they preserve half the supersymmetry of the fundamental  type
II string
theory.  This agrees with the fact that they are  BPS solitons of
the type II
theories.

The  effective low-energy action for superstring theory  in the
presence of a
solitonic
\RR $p$-brane can, in principle,  be written in the form
\begin{equation}\label{stot}
S = S_{bulk} + S^{p}_{source},
\end{equation}
where the bulk action is a ten-dimensional integral while the source is
restricted to the $p+1$-dimensional world-volume.   The equations
of motion
that follow from this
action arise in  the underlying string theory from the consistency
conditions
that ensure conformal invariance of the world-sheet theory (in the
case of the
type IIB theory there is no obvious covariant bulk action  but the
source-free equations   of motion are known \cite{schwarzsugra,howe}).
These bulk terms
arise from the usual closed-string sector of the theory but the
source terms
come from world-sheets with boundaries on which there can be a
condensate of
the open-string fields.   These source  actions, which  are of the
Born--Infeld
type,
reproduce the  equations of motion that follow from the consistency
conditions
for string theory in the background that includes an open-string
boundary
condensate, $F=dA$, of the electromagnetic field associated with
the abelian
vector potential that arises as a massless open-string state.  Such
consistency
conditions
on  open-string theory  were
originally derived in \cite{callana,polchinq,fradkina} in the case of the
type I
theory (which has Neumann boundary conditions in all ten space-time
dimensions).

The open-string world-sheet spanned by a  ground-state fluctuation of a
$D$-brane is a disk with the appropriate boundary conditions -- Neumann
in $p+1$ dimensions and Dirichlet in the remaining $9-p$.   This
world-sheet
can be
represented by a semi-infinite cylinder describing the evolution of
the vacuum
state at $\tau =-\infty$ to the  boundary (which we shall take to
be at $\tau
=0$).  With a closed-string state in the   \RR or \NSNS sector
coupling  to
the disk the semi-infinite cylinder describes the evolution of a
closed string
state $|\Phi\rangle$ from $\tau =- \infty$ to the boundary at $\tau =0$,
\begin{equation}\label{diskcoup}
\langle \Phi \rangle_{Disk} = \langle \Phi|B, \eta\rangle,
\end{equation}
where $|B,\eta\rangle$ denotes the boundary state and  $\eta = \pm
1$  labels
whether the state is BPS or anti-BPS.       In the formalism with
world-sheet
supersymmetry there are separate boundary states in each sector, $|B,
\eta\rangle_N$ and $|B, \eta\rangle_R$,  which  both satisfy the boundary
conditions
\begin{equation}\label{boses}
 (\partial X^i-\bar \partial X^i)|B,\eta\rangle =0, \qquad
(\partial X^\alpha
+ \bar\partial X^\alpha)|B,\eta\rangle=0,
\end{equation}
that impose Neumann conditions  on $p+1$ directions and Dirichlet
conditions on
the rest.       In order for the boundary state to preserve the
superconformal invariance of the bulk CFT, the super stress-energy
tensor has
to satisfy continuity conditions at the boundary, which take the form
\cite{polchina,greeng}
\begin{equation}\label{supergauge}
(F  + i\eta \tilde F) |B,\eta\rangle =0,
\end{equation}
where the generators of world-sheet supersymmetry transformations
are defined
by $F=\psi^\mu \partial X^\mu$ and  $\tilde F=\tilde{\psi}^\mu\bar{
\partial}
X^\mu$.  These conditions preserve half the world-sheet
supersymmetry.  The
boundary conditions on the world-sheet fermions follow from the
consistency of
(\ref{supergauge}) with (\ref{boses}),
\begin{equation}\label{ferm}
 (\psi^i + i\eta \tilde \psi^i)|B,\eta\rangle=0, \qquad
(\psi^\alpha- i\eta
\tilde \psi^\alpha )|B,\eta\rangle=0.
\end{equation}

 The fact that these states preserve half the world-sheet
supersymmetry is related to the fact that they also preserve half the
space-time  supersymmetry
which  relates the boundary  states in  the \RR and \NSNS sectors,
$|B,\eta\rangle_N$ and $|B,\eta\rangle_R$ \cite{lia}.

It is also possible to formulate the problem in a
light-cone gauge
\cite{greeng}, in which case space-time supersymmetry is manifest.  This
will be the
subject of this paper.  In the following
discussion the light-cone \lq time' coordinate  will be taken to
lie along the
axis of the cylinder which is one of the directions {\it
transverse} to the
brane.  In this parameterization $ X^+ = x^+ + p^+ \tau$  and the
momentum
component  $P^+$
is constant.   Since  the boundary state, $|B,\eta\rangle$,  is at
a fixed  \lq
time' $X^+$   satisfies a Dirichlet boundary condition.
Furthermore, $X^-$
is determined in terms of the transverse $X^I$ coordinates
($I=1,2,\cdots,8$)
and the world-sheet fermionic $SO(8)$ spinor coordinates $S^a$ and
$\tilde S^a$
  ($a=1,\cdots,8$) and also satisfies  a Dirichlet  boundary
condition as will
be seen in section 2.     This  means that there are at least two
Dirichlet
directions and that one of these is
time-like.  This kinematics describes a \lq (p+1)-instanton' rather
than a
$D$-brane (where time is one of the Neumann directions).  It is
related to the
$D$-brane by a double Wick rotation.   For convenience the words
\lq D-brane'
or \lq $p$-brane' will often be used in the following when it is really
this Wick
rotated version that is under consideration.   Since all cases will have
at least two
Dirichlet directions  the value of $p$ will be restricted to $-1\le
p\le 7$ in
the following.

The generic light-cone boundary state that preserves half of the 32
space-time
supersymmetry components of the type II theories   will be obtained
in section
2.  It is a generalization to $p>-1$ of the point-like state that
describes the
$D$-instanton \cite{greenc} and can be related to the $p=-1$ case by a
finite $SO(8)$
rotation.  Such a state has an interpretation   as  the top or
bottom state of
a light-cone superfield (the two cases corresponding to BPS and anti-BPS
states).   The general description of a boundary state must take
into account
the possibility of a non-trivial boundary condensate of the
open-string fields,
in particular the massless abelian gauge potential.  This
light-cone boundary
state has a form that automatically combines the \NSNS and \RR sectors.

A single boundary is interpreted as a linearized source of the closed-string
fields that
contributes to
$S_{source}$ in the effective action.  In section 3 it will be seen
that the
light-cone  description of the source  agrees, at linearized order,
with the
source terms that define the black $p$-brane solutions of the \RR sector
\cite{stromhorowitz,dufflub}.  In the presence of a boundary condensate the
effective source
actions are expected to be of the Born--Infeld
form \cite{schmidhubera,townsenda,dealwis,tseytlina}.   Expanding
these to linear
order in the
bulk fields also reproduces the light-cone frame boundary sources.

Configurations of two or more branes are considered in section 4.
The force between two $D$-branes is determined to lowest order by the
world-sheet that is a
cylinder with one boundary lying in the world-volume of each brane (this
generalizes the diagram for the $D$-instanton of
\cite{greeng,greenc,polchinc}).  The  force vanishes between parallel
identical branes  -- this is due to a cancellation between the
closed-string exchanges in the \NSNS and \RR sectors which is
characteristic of a BPS system that preserves 1/2 the total space-time
supersymetry.  More generally two
(non-identical) branes may be parallel and separated, they may
intersect (when they share at least one world-volume direction but are
not parallel) or
they may be orthogonal (when all the euclidean world-volume
directions are
orthogonal).  All possibilities are considered that are consistent
with the
choice of light-cone frame (the two world-volumes can occupy up to eight
dimensions transverse to the $x^\pm$ directions).    The number of
unbroken space-time supersymmetries may be 1/2 or 1/4 of the total as
is easily  deduced  by
considering a
$SO(8)$ rotation  that reduces any configuration to a standard one.
Configurations of two branes that preserve 1/4 of the total
supersymmetry again exert no force on each other but in this case the
force cancels separately within the \NSNS and \RR sectors.
Various configurations of type IIA and type IIB theories  with three
and four branes are also considered in section 4, in which
case the minimum unbroken supersymmetry may be $32/2^n$, where $n$ is
the number of branes.  One example is the intersection of three
three-branes with world-volumes that each share two common axes --
when compactified on $T^6$ this describes a black hole with
$a=0$. The results of this section are
light-cone versions of the description of intersecting branes given in
\cite{polchinskinotes}.

The  spontaneously broken space-time supersymmetries are considered
 in section
5.  These non-linearly realized symmetries relate S-matrix elements with
different numbers of soft fermions.

\section{Space-time Supersymmetry}

In the light-cone parameterization of the type IIB theory in which $X^+ = x^+ +
p^+ \tau$ the
coordinate $X^-$ is determined by the relation,
\begin{equation}\label{xmindef}
p^+ \partial_\sigma X^- = \partial_\tau X^I \partial_\sigma X^I - i S^a
\partial_{\tau-\sigma} S^a
+  i\tilde S^a \partial_{\tau  +  \sigma} \tilde S^a
\end{equation}
(whereas the right-moving spinor in the type
IIA theory
is a dotted $SO(8)$ spinor). We can anticipate that the boundary
conditions on
the  fermionic coordinates will reflect a $S$ into $iM\tilde S$
where $M$ is an
orthogonal matrix so that the fermionic terms will cancel in the
expression for
 $\partial_\sigma X^- |B,\eta\rangle$.  It follows that $X^-$
satisfies the
Dirichlet condition,
\begin{equation}\label{dirxmin}
\partial_\sigma X^-=0,
\end{equation}
whether the $X^I$ are Dirichlet or Neumann coordinates.

The  coordinates transverse to the $\pm$ directions  in
(\ref{boses}) satisfy
the boundary conditions
\begin{equation}\label{bosboun}
(\partial X^I - M^{p I}_{\ \ J} \bar \partial X^J) |B,\eta\rangle_{(p)}
=0,\end{equation}
where $M^p_{IJ}$ is an element of  $SO(8)$.  The Neumann directions
will be
chosen to be $\alpha = I =1, \cdots, p+1$ while the Dirichlet
directions will
be chosen to be $i = I = p+2, \cdots ,8$ (the superscript and subscript
${p}$ will often be dropped in the following when there is no
ambiguity).    In
the absence of a boundary
condensate  of the massless  open-string vector potential $M_{IJ}$  can be
written in block diagonal form,
\begin{equation}\label{mdeff}
M_{IJ} = \pmatrix {- I_{p+1}& 0 \cr
      0 &   I_{7-p}  \cr}
\end{equation}
(where $I_q$ indicates the $(q\times q)$-dimensional unit matrix).
In the more general case in which there is a boundary condensate of the
open-string vector
potential  (to be considered in the next section)   $M_{IJ}$  is a more general
$SO(8)$ matrix and can be  written as,
\begin{equation}\label{mgeneral}
M_{IJ} =   \exp\left\{\Omega_{KL}  \Sigma^{KL}_{IJ} \right\},
\end{equation}
where   $\Sigma^{KL}_{IJ} =  (\delta^K_{\ I}\delta^L_{\ J} - \delta^L_{\
I}\delta^K_{\ J}) $ are generators of $SO(8)$ transformations in
the vector
representation.  The parameters  $\Omega_{IJ}$ depend on the open-string
boundary condensate and  in a
particular basis they  can be written in block off-diagonal form as,
\begin{equation}\label{omdef}
\Omega_{\alpha\beta} =   {\rm diag}(D_1, D_2, \cdots, D_{(p-1)/2}),
\end{equation}
with
\begin{equation}\label{matmid}
D_m = \pmatrix{0& -d_m\cr
       d_m & 0 },
\end{equation}
while the components $\Omega_{ij}$ and $\Omega_{\alpha i} $   may be
taken to vanish
for a static $p$-brane \cite{bachasb,callankleb}.  In the
absence of a condensate  $d_m = \pi$.  In other words the Neumann
boundary
conditions in $p+1$ directions are represented by a rotation of $\pm\pi$
transverse to $(p+1)/2$ axes.   In the presence of a condensate
of the open-string vector  potential  the rotations differ from
$\pi$.

Each of the  sixteen-component supercharges of the type II theories
decompose
in the light-cone treatment into two inequivalent  $SO(8)$ spinors
defined in
terms of the world-sheet fields $X^I$  and $S^a$ ($a=1,\cdots,8$) by,
\begin{equation}\label{susyalg1}
Q^a ={1\over \sqrt{2p^+}}\int_0^{\pi}d\sigma S^a(\sigma), \qquad
 {Q}^{\dot a}=  {1\over \pi \sqrt{p^+}}\int_0^{\pi }d\sigma
\gamma^I_{\dot a b}
 \partial X^I
 {S}^b(\sigma),
\end{equation}
for the left-moving charges and similar expressions for the right-moving
charges, $\tilde Q^a$ and $\tilde Q^{\dot a}$ expressed in terms of the
right-moving coordinates.   The undotted supercharge acts linearly
while the
dotted supercharge acts non-linearly on the world-sheet fields (and
$\gamma^I_{a\dot a}$ are the usual $SO(8)$ gamma matrices).
These charges
realize  the $N=2$ supersymmetry algebra in the light-cone gauge
\begin{equation}\label{algsus}
\{Q^a,Q^b\}= 2p^+ \delta^{ab},\quad \{Q^{\dot{a}},Q^{\dot{b}}\} =
\delta^{\dot{a}\dot{b}} P^- , \quad \{Q^a,Q^{\dot{a}}\} ={1\over
\sqrt 2}
\gamma^I_{a\dot{a}}p^I,
\end{equation}
with a similar algebra for the rightmoving charges.  The closed-string
light-cone hamiltonian, $P^-_{cl}$, is defined by
$P^-_{cl} = P^- + \tilde P^-$, where
\begin{equation}\label{pmindef}
P^- =  {1 \over 4 p^+} (p^\alpha)^2 + {1\over 2 p^+} \sum_{n=
1}^\infty\left(
\alpha^I_{-n} \alpha_n^I  +   n  S^a_{-n} S_n^a  \right),
\end{equation}
and the modes of $X$, $S$ and $\tilde S$ are defined in the usual manner.

In the type IIB theory both the left-moving and right-moving
linearly realized
supercharges are undotted spinors whereas in the type IIA theory
the dotted and
undotted indices are switched between the left-moving and
right-moving charges.
The case of the type IIB theory will be described first.

The boundary state is defined to  conserve the linear combinations of
space-time
supercharges,
\begin{eqnarray}\label{susyi}
Q^{+a}_\eta |B,\eta\rangle\equiv (Q^a + i\eta M_{ab} \tilde
Q^b)|B,\eta\rangle
=0, \nonumber \\
Q^{+\dot a}_\eta |B,\eta\rangle \equiv  (Q^{\dot a} + i\eta M_{\dot
a \dot b}
\tilde Q^{\dot b})|B,\eta\rangle =0,
\end{eqnarray}
which generalize the expressions  in \cite{greeng}   which applied to the
special
case of the $D$-instanton( for which $M_{ab}=\delta_{ab}$, $M_{\dot
a\dot b} =
\delta_{\dot a\dot b}$).   The remaining  combinations
\begin{equation}\label{mindef}
Q^{-a}_\eta \equiv (Q^a  -  i\eta M_{ab} \tilde Q^b) ,\qquad
Q^{-\dot a}_\eta
\equiv  (Q^{\dot a} -  i\eta M_{\dot a \dot b} \tilde Q^{\dot b})
\end{equation}
are the broken supersymmetries that are associated with goldstinos
as will be
described later.

In order to determine $M_{ab}$ and $M_{\dot a\dot b}$  we first
make the ansatz
that
the   $SO(8)$ fermionic world-sheet fields,  $S^a$ and $S^{\dot
a}$, satisfy
the boundary conditions,
\begin{equation}\label{modsusy}
(S_n^a+i\eta M_{ab}\tilde{S}^b_{-n})|  B, \eta \rangle =  0, \qquad
(S_n^{\dot{a}}+i \eta M_{\dot{a}\dot{b}}\tilde{S}^{\dot{b}}_{-n})|  B,
\eta\rangle = 0.
\end{equation}
The  bispinor matrices  $M_{ab}$ and $M_{\dot a\dot b}$ are now
determined by
consistency with the superalgebra.   Thus, multiplying the first
equation in
(\ref{modsusy}) by $(S^a_{-n} + i\eta M_{ab} \tilde S^b_n)$
determines that $M$
is orthogonal,
\begin{equation}\label{cond1}
(M^T)_{ab}M_{bc}= \delta_{ac}.
\end{equation}
Furthermore, multiplying the second equation in (\ref{modsusy}) by
$(S^{\dot
a}_{-n} + i\eta M_{\dot a\dot b} \tilde S^{\dot b}_n)$ gives the
condition
(using  (\ref{bosboun}) and the definition of the nonlinearly realized
supercharges),
 \begin{equation}\label{cond2}
\gamma^I_{\dot a a} M^{IJ} -M_{ab}M_{\dot a \dot b} \gamma^J_{\dot
b b}=0.
\end{equation}
These two conditions are solved by $SO(8)$ rotations acting on the
spinors,
\begin{equation}\label{mdef}
M_{ab} = \exp\{\half \Omega_{IJ}\gamma^{IJ}_{ab}\}, \qquad \qquad
M_{\dot a\dot
b} = \exp\{\half \Omega_{IJ}\gamma^{IJ}_{\dot a \dot b}\},
\end{equation}
where $\Omega_{IJ}$ is the same antisymmetric matrix (\ref{omdef})  that
defined the $SO(8)$
rotation in the vector basis and $\gamma^{IJ} = \half (\gamma^I\gamma^J -
\gamma^J\gamma^I)$.   In the absence of a boundary condensate these
conditions reduce to
\begin{equation}\label{mdefte}
M_{ab} = \left(\gamma^1\gamma^2 \dots
\gamma^{p+1}\right)_{ab},\qquad M_{\dot
a\dot b} = \left(\gamma^1\gamma^2 \dots \gamma^{p+1}\right)_{\dot
a\dot b}.
\end{equation}
The three matrices $M_{IJ}$, $M_{ab}$ and $M_{\dot a\dot b}$ are
related to
each other by triality.

The boundary state that solves (\ref{susyi}) can now be obtained
explicitly
as,
\begin{eqnarray}\label{boundarystate2}
|  B\rangle & =\exp\sum_{n>0}\left(   {1\over n}  M_{IJ}
\alpha^I_{-n}\tilde{\alpha}^J_{-n} -
i  M_{ab}S^a_{-n}\tilde{S}^b_{-n}\right)|B_0\rangle,\nonumber \\
&= R(M) \exp \sum_{n>0} \left({1\over n}
\alpha^I_{-n}\tilde{\alpha}^I_{-n}  -
i   S^a_{-n}\tilde{S}^a_{-n}\right)|B_0\rangle,
\end{eqnarray}
where the zero-mode factor is
\begin{equation}\label{boundarystate1}
|  B_0\rangle =C\left(   M_{IJ} |  I\rangle | J\rangle +i
M_{\dot{a}\dot{b}} |
\dot{a}\rangle| \dot{b}\rangle \right)
\end{equation}
(and is annihilated by all the positive modes).
The argument $\eta$ has been dropped  since a state with one value
of $\eta$
(an anti-BPS state, say)  can be transformed into a state with the
opposite
value (a BPS state) by a $2\pi$ rotation about all axes -- this
leaves $M_{IJ}$
unchanged but reverses the sign of $M_{ab}$ and $M_{\dot a\dot b}$.   In
verifying that this state satisfies (\ref{susyi}) use
is made of the relations
$S_0^a|  I\rangle=  \gamma^I_{a\dot{a}} |\dot{a}\rangle/\sqrt 2$
and $S_0^a
|\dot{a}\rangle =  \gamma^I_{a\dot{a}} |I\rangle/\sqrt 2$.
The normalization constant $C$  can be determined by  constructing a
cylindrical world-sheet  by joining two boundaries together with a
closed-string propagator.  The cylinder is equivalent to an annulus
that can be
uniquely determined as a   trace over open-string states  with
end-points fixed
on the respective branes.  In the absence of a condensate of the
open-string
field on the boundaries $C=1$ but, more generally, its value
depends on the
boundary condensate  \cite{polchinq,callan3}.

The analogous boundary  state for the type IIA theory involves a matrix
$M_{a\dot b}$
that is the product of an odd number of $\gamma$ matrices.

The operator  $R(M)$ in (\ref{boundarystate2}) is the representation of
$SO(8)$ rotations on
the non-zero
modes and is defined by
\begin{equation}\label{rdef}
R(M)=\exp  \sum_{n>0} \left({1\over n}  T^{(\alpha)}_{IJ}
\alpha^I_{-n}{\alpha}^J_{n}   +   T^{(S)}_{ab}
S^a_{-n}{S}^b_{n}\right) , \end{equation}
where
\begin{eqnarray}\label{gendefs}
T^{(\alpha)}_{IJ} = \Omega_{KL} \Sigma^{KL}_{IJ} , \qquad
T^{(S)}_{ab} = {1\over 2}  \Omega_{KL} \gamma^{KL}_{ab}.
\end{eqnarray}
This satisfies  the group property $R(M_1) R(M_2) = R(M_1M_2)$.
Similarly,
apart from the overall scale $C$, the zero-mode part of the state
(\ref{boundarystate1}) is a rotation of ground-state scalars, which
can be
written symbolically as
\begin{equation}
|  B_0\rangle_{(p)} = C R_0(M^p) \left( |  I\rangle | I\rangle +i
|\dot{a}\rangle| \dot{a}\rangle \right) = CR_0(M^p) |B_0\rangle_{(-1)} .
\end{equation}

Thus all the $D$-brane boundary states are obtained, up to a
normalization, by
 $SO(8)$ rotations of the boundary state of the $D$-instanton  (the
purely Dirichlet case, $p=-1$),
\begin{equation}\label{rotins}
|B\rangle_{(p)} = C\hat R(M^{p}) |B\rangle_{(-1)},
\end{equation}
where $\hat R = R_0 R$.   The cases with $p>-1$ are obtained by
rotations
through $\pi$ around $(p+1)/2$ axes while non-zero condensates of the
open-string vector potential  are determined by continuous rotations.  A
rotation of $2\pi$ around all the axes changes a
BPS state into an anti-BPS state.

\vskip 0.3cm
\noindent{\it Light-cone superfields}
\vskip 0,2cm

The boundary state can also be expressed as a light-cone superfield by
introducing Grassmann coordinates defined by,
\begin{equation}\label{grass2}
\theta^a = \half (Q^a-i\tilde{Q}^a)
\end{equation}
which is conjugate to
\begin{equation}\label{grass1}
\frac{\partial}{\partial\theta^a} = {1\over 2p^+} (Q^a+i\tilde{Q}^a) .
\end{equation}
The bosonic sector of the closed-string superfield boundary state
can then be
expressed as a power series in
the $\theta^a$   by writing the zero-mode  factor as,
\begin{equation}\label{superf}
|B_0, \theta ,\eta\rangle= \sum_{N=0}^4 {1\over (2N)!} (ip^+)^{N-2}
A^{2N}_{a_1a_2\dots a_{2N}}\theta^{a_1}
\theta^{a_2}\dots \theta^{a_{2N}}   |  0,\eta\rangle.
\end{equation}
The coefficients in this expansion are complex functions that satisfy the
constraint implied by the \lq reality' condition,  $A^p =
({A^{8-p}})^*$ so that
there are $2^8$ real bosonic states.

The supersymmetry conditions  (\ref{susyi})  restrict the boundary
states so
that in the case of the $D$-instanton ($p=-1$)  the BPS and
anti-BPS states
satisfy  the superspace conditions
 \begin{equation}\label{newboun}
\frac{\partial}{\partial\theta^a}| B, \theta, + \rangle  =0,
\qquad  \theta^a|
B, \theta,  - \rangle  =0,
\end{equation}
which means  that the boundary states are the top or bottom
components of the
superfield,  (\ref{superf}) depending on whether they are BPS or
anti-BPS.

With $p>-1$  the conditions (\ref{susyi}) can be satisfied by
requiring linear
combinations of $\theta$ and $\partial/\partial\theta$ to
annihilate the state.
 This is conveniently expressed in terms of a modified Grassmann
coordinate,
\begin{equation}\label{newthet}
\hat \theta = {1\over 2}(1 + M)_{ab}\theta^b + {p^+ \over 2} (1-
M)_{ab}{\partial\over
\partial \theta^b}
\end{equation}
and its conjugate
\begin{equation}\label{newpart}
{\partial \over \partial \hat \theta^a} = {1\over 2p^+}  (1-M)_{ab}
\theta^b + {1\over 2}(1
+ M)_{ab}{\partial \over \partial \theta^b} .
\end{equation}
The components of the  superfield that is a function of $\hat \theta'$
are linear
combinations of the components of the field (\ref{superf}).
The conditions (\ref{susyi}) are satisfied in general by
\begin{equation}\label{newsols}
\frac{\partial}{\partial\hat \theta^a}| B_0, \hat \theta, + \rangle =0,
\qquad
\hat  \theta^a| B_0,\hat  \theta, - \rangle=0.
\end{equation}
This means that  the boundary state is the top or bottom component of the
light-cone superfield defined as an expansion in $\hat \theta$.
This applies
 for any value of $p$.

The complete boundary state may be expressed  in a string superspace
by introducing a a Grassmann $SO(8)$   world-sheet spinor coordinate,
$\hat \Theta^a$ (with zero mode $\hat \theta^a$), and defining a
string superfield $\Phi[X^I,
\hat \Theta^a,  p^+, x^+] = \langle X^I,
\Theta^a, p^+, x^+| \Phi\rangle$, where $|\Phi\rangle$ denotes a
general closed-string state.   The string field theory source term takes the
form,
\begin{equation}\label{sourcestr}
 \langle \Phi | B\rangle = \int dx^+ dp^+ D^8X^I D^8 \hat \Theta^a
\delta^{7-p}(X^i) \delta^8(\hat \Theta) \delta(x^+)
\Phi
\end{equation}
($ i =p+2,\cdots,8$) where $\int D^8 \hat \Theta \delta^8(\hat \Theta)$ picks
out the bottom state in the (infinite-dimensional) stringy light-cone
supermultiplet, which is the BPS state.  The source for the anti-BPS state
would not contain the factor $\delta^8(\hat \Theta)$.   The full light-cone
string field theory action in the presence of the BPS source term has the form,
\begin{equation}\label{fieldlag}
S= \langle \Phi | 2p^+  (P^- - p^-)| \Phi\rangle  + \langle \Phi |
B\rangle + {\rm interaction \ terms},
\end{equation}
where the integration over the super space-time coordinates, $X^I$, $\hat
\Theta$, $p^+$
and $x^+$ is implied by the notation \cite{greenvacuum}.

\section{Source equations}

The zero-mode factor in the boundary state  determines the coupling
of the
$D$-brane to the massless fields so the relative strength of the
source terms
in the effective string action $S_{source}$ in (\ref{stot}) can be
seen  from
(\ref{boundarystate1}).  Decomposing
$M_{IJ}$ into $SO(8)$ representations $8\otimes 8 = 35 + 28 +1$,
\begin{equation}\label{nsnsst}
M_{IJ} = M_{(IJ)} + M_{[IJ]} + {1\over 8} \eta_{IJ} M_{KK}
\end{equation}
where $(\ ) $ denotes the symmetric traceless part and $[\ ]$ denotes the
antisymmetric part.   These terms couple to the transverse
graviton,  $G_{IJ}$,
 and  antisymmetric tensor, $B^N_{IJ}$, as well as to the  dilaton
of the type
II theories.  The \RR component in the boundary state can be
written as the sum
of $SO(8)$ representations,
\begin{equation}\label{spins}
M_{\dot a \dot b} = {1\over 8} \delta_{\dot a \dot b} \tr M + {1\over 16}
\gamma^{IJ}_{\dot a \dot b} \tr (\gamma^{IJ} M) + {1\over 384}
\gamma^{IJKL}_{\dot a \dot b} \tr(\gamma^{IJKL}M),
\end{equation}
which defines the couplings of the boundary state to the transverse
components
of the massless \RR potentials,  $\chi$, $B^R_{IJ}$ and
$A^{(4)}_{IJKL}$ (which
is self-dual in the transverse space).   The information in the
boundary state
thus determines the relative strengths of  the sources for the
fields in the
theory that modify the bulk field equations.

Covariant string perturbation theory takes place in the linearized
Siegel gauge,

\begin{equation}\label{segalga}
\partial_\nu h^\nu_{\ \mu} -{1 \over 2} \partial_\mu h^\nu_{\ \nu}  + 2
\partial_\mu \phi  =0,
\end{equation}
where  $h_{\mu\nu} = G_{\mu\nu} - \eta_{\mu\nu}$ is the metric
fluctuation.
The light-cone gauge is obtained by setting $h^{+\nu}=0$.   The
light-cone
gauge graviton is then identified with the traceless part of the metric
fluctuation, $\hat h_{IJ} = h_{IJ} -\eta_{IJ} h^K_{\ K}/8$ and the dilaton is
proportional to the trace of the metric fluctuation,
\begin{equation}\label{dilatr}
\hat \phi= \phi -\phi_0  = {1\over 4} h^I_{\ I} = {1\over 4} (h^i_{\ i} +
h^\alpha_{\ \alpha}).
\end{equation}

In the space-time supersymmetric light-cone gauge the fluctuations of all the
fields are
normalized in a
manner that is independent of the coupling constant, $g=e^{\phi_0}$
However, in the covariant
action the  terms in the \NSNS sector have a  normalization factor
of $1/g^2$.
This means that in making a comparison between the light-cone gauge
results and
the expansion of the covariant action in small fluctuations it will be
necessary to redefine the \NSNS fields by a factor of $g$.  In
other words,
denoting the covariant field fluctuations  by a tilde, we shall make the
identification $h_{IJ} = \tilde h_{IJ}/g$  for the metric
fluctuation, $\phi =
\tilde \phi/g$  for the dilaton  and $B^N_{IJ} = \tilde B^N_{IJ}/g$
for the
antisymmetric tensor.

\subsection{Black $p$-brane Ans\"atze}

 The values of the boundary sources can be compared directly with
the explicit
$p$-brane solutions \cite{stromhorowitz} of the effective type II
supergravity
theory.   In the absence of a condensate of the open-string vector
potential
the couplings of the \NSNS fields are determined by the symmetric matrix,
$M_{IJ} = {\rm diag} (- I_{p+1},  I_{7-p})$.
In the light-cone frame the graviton $\hat h_{IJ}$  couples to the
traceless
part of this matrix,
\begin{equation}\label{lccoms}
 {\rm diag} ( - {1\over 4}(7-p) I_{p+1},   {1\over 4}(p+1) I_{7-p}),
\end{equation}
while  the dilaton coupling is proportional to the trace,
\begin{equation}\label{dilcoup}
{1\over 4} \tr M^I_{\ I} = -{1\over 2}(p-3).
\end{equation}
Similarly, the \RR $(p+1)$-form couples to couplings are proportional to
$M_{ab} = (\gamma^1\cdots \gamma^{p+1})_{ab}$.   If  the \lq
electric' charge
is carried by an object with a $(p+1)$-dimensional world-volume the \lq
magnetic' charges are carried by objects with a $(p'+1)$-dimensional
world-volume where $M^{p'}_{ab} = * M^{p}_{ab}$, where $*$ denotes
the Hodge
dual in the eight-dimensional transverse space, $M^{p}= \gamma^1 \cdots
\gamma^{7-p}$.

The couplings of the $D$-brane sources to the dilaton, graviton and
\RR potentials determine the source terms in the effective field
theory in linearized approximation around flat space.  These terms can
be compared with those associated with the\RR sector  black
$p$-brane solutions   of \cite{stromhorowitz}.  These are
solutions of the
type II supergravity theories that preserve one half of the space-time
supersymmetries in which the metric and the dilaton take the
following form in
the string frame,
\begin{eqnarray}
ds^2&=& A^{-1/2} (x^i)dx^\alpha
dx^\beta\eta_{\alpha\beta}+A^{1/2}(x^i)dx^i
dx^j
\eta_{ij}\nonumber \\
e^{\phi}&=&A^{-(p-3)/4}\label{blackbra}
\end{eqnarray}
Here, the function $A$ depends only on the transverse coordinates
$x^i$ ($i=p+2, \cdots, 7 - p$) and satisfies $\partial^2 A=0$ for
$x^i\neq 0$.
It is given by,
\begin{equation}
A(x^i)=1+\frac{Q}{\mid x^i\mid^{7-p}}
{}.
\end{equation}
where $Q$ is a (quantized) charge.

To make the comparison with the light-cone boundary state the
solution should
first be Wick rotated so that the world-volume has euclidean
signature and the
transverse space has  lorentzian signature.  The solution
(\ref{blackbra})  can
then be expanded  in small fluctuations around flat space as $|x^i|
\to \infty$
(the space-times of black p-branes are asymptotically minkowskian),
using
\begin{equation}
A^a=(1+\frac{Q}{\mid x^i\mid^{7-p}})^a\sim 1+a\frac{Q}{\mid
  x^i\mid^{7-p}}+ O(\mid x^i\mid^{-14+2p}).
\end{equation}
The second term corresponds to small fluctuations around
Minkowski space and is generated by a source of strength
proportional to $a$.
After transforming to the light-cone gauge the metric fluctuations are
non-trivial in the transverse space with a  trace that is
identified with the
dilaton while the traceless part determines the physical graviton.  The
coefficients of these terms agree with those obtained from the massless
$D$-brane sources.
The correspondence between D-branes and the black p-brane solutions
can also be analyzed by comparing scattering  massless closed string
states  of D-branes and scattering in the black p-brane background
\cite{klebanovgravlensing,myers}.

\subsection{ Boundary condensate}

The boundary action that describes the  coupling of a condensate of the
massless open-string gauge potential has the form
\begin{equation}\label{bouncons}
\int ds \left(  A_\alpha   \dot X^\alpha  -  {i\over 2}  F_{\alpha\beta}
S\gamma^{\alpha\beta} S \right).
\end{equation}
where $s$ is the parameter on the world-sheet boundary and
$F_{\alpha\beta} =
\partial_{[\alpha} A_{\beta]}$.  In addition,  if $B^N_{\alpha\beta}$ is
constant the Wess--Zumino term in the bulk action is a total
derivative and
may be expressed as a surface term so that the total surface term is
\begin{equation}\label{bountot}
S_{surface} =     \int ds \calF_{\alpha\beta} \left( X^{[\alpha}
\partial_s
X^{\beta]} - i S \gamma^{\alpha\beta} S\right),
\end{equation}
where $\calF = F-B^N$.  This mixing of the closed-string
antisymmetric tensor
and the open-string vector was considered in  \cite{Cremmerscherk}.
The quantity in parentheses in this expression
is the generator of rotations in the $\alpha-\beta$ direction.
This condensate
can be expected to be a consistent string background if the field
strength is slowly varying (so that terms involving the derivative of
$\cal F$ may be dropped), which is all that will be considered here.
The presence
of this
boundary term  in the action leads to a modification of the
boundary conditions
in the Neumann directions, giving  $(\partial_n X^\alpha +
\calF_{\alpha\beta}
\partial_t X^\beta)|B,\eta\rangle =0$ where $n$ and $t$ are the
normal and
tangential directions to the boundary.  These conditions are
equivalent to
(\ref{rdef}) with
\begin{equation}\label{bounstate}
M^p_{\alpha\beta} = - \left[(1-\calF)(1+\calF)^{-1}\right]_{\alpha\beta}.
\end{equation}
and the normalization constant $C$ in the end-state $|B_0\rangle$ may
be determined to be,
\begin{equation}
C=\sqrt{\det(1+\calF)}.
\end{equation}

\vskip 0.3cm
{\noindent{\it\NSNS  sources and Born-Infeld actions}}
\vskip 0.2cm
The  boundary source terms lead to modifications of the effective
low-energy
field theory as was studied in the purely Neumann type I theory in
\cite{callana,fradkina}.
There, BRST invariance in the presence of boundaries was shown to
lead to an
effective action that includes a source term, $S_{source}$, that has a
dependence on the open-string massless vector potential that is of the
Born--Infeld form.

In the context of a  $p$-brane similar arguments again lead to a
Born--Infeld-like source action \cite{leigha}.   This is  defined in a
$(p+1)$-dimensional space which has euclidean signature when the
kinematics are
appropriate to our transverse light-cone gauge.
The \NSNS  part of this action  has the  form,
\begin{equation}\label{effact}
S^p_{NS} = \int d^{p+1}x  e^{-\phi} \sqrt{\det (G+{\cal F})} .
\end{equation}
This can be compared with the source terms that arise from the massless
components in the boundary state  (\ref{boundarystate1}) by expanding
(\ref{effact}) in small fluctuations of the bulk
closed-string  fields around their constant background values.  These
fluctuations comprise the dilaton  $\tilde \phi = \phi - \phi_0$
and the metric
and antisymmetric tensor fields that may be combined into
$\tilde h_{\alpha\beta} = \eta_{\alpha\beta} - G_{\alpha\beta} +
B_{\alpha\beta}^N$.
Recalling that $\tilde \phi =
 (\tilde h^i_{\ i} + \tilde h^\alpha_{\ \alpha})/4$ the expansion of the
determinant factor in (\ref{effact}) gives
\begin{eqnarray}\label{detfact}
&&-{1\over g} \tilde \phi \sqrt{\det(1+\calF)}  + {1\over 2g}
\sqrt{\det(1+\calF)} \tilde h^{\alpha\beta}
(1+\calF)^{-1}_{\alpha\beta} \nonumber\\
&& \qquad\qquad =
{1\over 4g} \sqrt{\det(1+\calF)}  \left(-\tilde h_{ii} +
\left({1-\calF\over
1+\calF}\right)
_{\alpha\beta} \tilde h^{\alpha\beta}\right),
\end{eqnarray}
where $g= e^{\phi_0}$.  Taking into account the fact that the light-cone
graviton is defined by $h_{IJ} =  \tilde h_{IJ}/g$,
this  is precisely the same as the source obtained from the massless
part of the
boundary state
(\ref{boundarystate1}).

\vskip 0.3cm
\noindent{\it Boundary condensate and R-R fields}
\vskip 0.2cm

The couplings between the boundary field strength  $\calF$ and the \RR
sector massless
potentials of the type IIB theory are simplest to express in the frame in which
$\calF$ is block
off-diagonal,
\begin{equation}
{\cal F}_{IJ}={\rm
diag}({\cal{F}}_{1},\cdots,{\cal{F}}_{(p+1)/2},0,\cdots,0)
\end{equation}
where ${\cal{F}}_{i}$ is given by
\begin{equation}
{\cal{F}}_{i}= \pmatrix{ 0& -f_i\cr
f_i&0\cr}
\end{equation}
(for $p$ odd).
The matrix $M_{ab}$ which determines the coupling to the $RR$ states
is then given by the following expression
\begin{equation}\label{fidef}
M_{ab}=\prod_{i=1}^{(p+1)/2} {1 \over (1+ f_i^2)^{1/2}}
(1+f_i\gamma^{2i-1\
2i})\gamma^1\cdots\gamma^{p+1}
\end{equation}
The  contributions to the covariant  world-volume source action are determined
by this
expression.
 They are of Chern-Simons type, $\int d^{p+1}x  A_R^{k}(\wedge F)^n $,
where $A_R^k$ is a $k$-form potential in the \RR sector (and
$k+2n=p+1$).
For example, the matrices $M^p_{ab}$ for the cases $p=1$ and $p=3$  are given
by
\begin{eqnarray}
M^1_{ab} & = & {1 \over (1+ f_1^2)^{1/2}} \left(
\gamma^{12}_{ab}+f_1\delta_{ab}\right) \label{source1}\\
M^3_{ab} & = & {1 \over (1+ f_1^2)^{1/2} (1+ f_2^2)^{1/2}}
\left(\gamma^{1234}_{ab}+
f_2\gamma^{12}_{ab}+f_1\gamma^{34}_{ab}+f_1f_2\delta_{ab}\right).
\label{source2}\end{eqnarray}
Correspondingly,  covariant effective euclidean source actions for  these
fields are given by
\begin{eqnarray}
S^1_{R}& = & i\int d^2x (B^R+\chi {\cal F})\label{sonee}\\
S^3_{R} & = & i\int d^4x (\frac{1}{2}\chi {\cal{F}}\wedge{\cal F} +
B^R\wedge{\cal F} + A^4_R)\label{stwoe}.
\end{eqnarray}
Obvious generalizations of these expressions hold for all other values of
$p\le7$, including the even values of relevance to the type IIA theory.

Expanding these actions to linearized order in the bulk   fields
gives sources
for the \RR fields that correspond to the terms in
(\ref{source1}),(\ref{source2}).
   These terms also contain
couplings between
the bulk fields, such as $\chi \wedge B^N$ in  (\ref{sonee}) and
$B^N\wedge
B^N$, $B^R\wedge B^N$ in (\ref{stwoe}).  The presence of such
couplings can be
demonstrated directly by calculating the disk amplitude with the
appropriate
insertion of closed-string vertex operators.

\subsection{Examples}

\noindent{\it D-instanton}
\vskip 0.2cm

In the case  $p=-1$ all  coordinates
satisfy Dirichlet boundary conditions and
$M_{IJ}=\delta_{IJ
}$,  $M_{ab}=  \delta_{ab}$.  The  massless
terms in the
boundary state are \begin{equation} \label{dinstaton1}
|  B_0\rangle = \delta_{IJ}|  I\rangle\tilde{|
J \rangle} + i \delta_{\dot{a}\dot{b}}|  \dot{a}\rangle\tilde{|
\dot{b}\rangle}. \end{equation}
Thus, the couplings of the source to the dilaton $\phi$ and the R-R
scalar,
$\chi$, are equal which agrees with the linearized approximation to the
$D$-instanton ansatz in \cite{gibbonsa}.   The factor of $i$  indicates
that this
should be interpreted as a euclidean solution since the
pseudoscalar field
$\chi$ is pure imaginary  after a Wick rotation to  euclidean space.

The full theory includes a sum over diagrams with arbitrary numbers of
world-sheet boundaries.
The leading term in the partition function comes from the
exponential of the
disk diagram so that
\begin{equation}\label{diskful}
Z = e^{-{1 \over g} - i \chi_0}.
\end{equation}
  Connected diagrams with a larger number of boundaries vanish by
supersymmetry.  The exponent in $Z$  is the
constant part
of the euclidean source action,
\begin{equation}\label{sourceford}
S^{-1}_{source}=\int d^{10}x(e^{-\phi}+i \chi)\delta^{10}(x-x_0).
\end{equation}
The fact that the $D$-instanton has no world-volume dimensions
means that
there is no dependence on the metric in the source action and hence
no coupling
to the graviton -- it couples to the dilaton and $\chi$ only.   The
source term
is therefore unchanged in transforming between the string and the
Einstein
frames.

\vskip 0.3cm
\noindent{\it The $D$-string}
\vskip 0.2cm

In the absence of a boundary condensate of  the open-string vector
potential
the $p=1$ boundary state couples to the
dilaton, the graviton, and the components of the \RR-antisymmetric tensor
$B^R_{12}$   in the world-sheet directions.

In the presence of a boundary condensate of  the open-string vector
potential the $D$-string also  couples   to the \RR scalar, $\chi$,
and the
\NSNS antisymmetric tensor, $B^N$ \cite{wittenb}.
This is seen in the present light-cone formalism by constructing
the boundary
state (\ref{bounstate}) for the case $p=1$, in which the rotation
matrices have the form (which is analogous to the covariant
description in \cite{lia})
\begin{equation}\label{ramthree}
M^1_{ab} = \frac{1}{\sqrt{1+f^2}}(\delta_{ab}+ f\gamma^{12}_{ab}),
\end{equation}
and
\begin{equation}\label{twobr}
M^1_{IJ} = \pmatrix{M^1_{\alpha\beta} (f)& 0 \cr
    0 & I_6\cr},
\end{equation}
where the $2\times 2$ matrix $M^1_{\alpha\beta}(f)$ is given by
\begin{equation}\label{twodef}
M^1_{\alpha\beta} = -  {1-\calF\over 1+\calF} = {-1\over
1+f^2}\pmatrix{1-f^2 &
 -2f \cr
     2f & 1-f^2 \cr}.
\end{equation}
where $f =-\half \epsilon^{\alpha\beta} \calF_{\alpha\beta}$.

The  $SO(8)$ rotation is through an angle $\alpha$ in the $0-1$
plane given by,
\begin{equation}\label{cosalph}
\cos{\alpha}= - \frac{1-f^2}{1+f^2},
\end{equation}
so that the generators of the rotations in (\ref{gendefs}) are
\begin{equation}\label{tdefss}
T^{(\alpha)}_{\alpha\beta} = \alpha \Sigma^{12}_{\alpha\beta}, \qquad
T^{(S)}_{ab} = {\alpha\over 2} \gamma^{12}_{ab}.
\end{equation}

The zero mode part of the boundary state is given by
\begin{equation}
|  B_0\rangle_{(1)} =  \sqrt{1+f^2}M_{IJ}|  I\rangle|
\tilde{J}\rangle+i(f\delta_{\dot{a}\dot{b}}+\gamma^{12}_{
\dot{a}\dot{b}})|  \dot{a}\rangle|  \tilde{\dot{b}}\rangle
\end{equation}
Following the general analysis of section 3.2 the effective
world-volume action of the D-string is given by  a Born-Infeld-like
euclidean action
\begin{equation}
S^1_{source} = \int d^2x(e^{-\phi}\sqrt{\det(G+\calF)}+i \calF\chi+ i
B^{R}_{12}),
\end{equation}
which reproduces the boundary sources in linearized approximation,
as before.

In order to make contact with the dyonic string solutions of
\cite{schwarzx} the action may be expressed in terms of the integer
monopole number of the integrated field strength.  This may be
accomplished by integrating over the vector potentials in a particular
topological sector in which $\int {\cal F} = 2\pi m$
\cite{schmidhubera,dealwis,tseytlina}.  The result is an
action of the Nambu--Goto form,
\begin{equation}\label{onesource}
\tilde{S}^1_{source}(m) = \int d^2x \left(
\left(e^{-2\phi} + (m+\chi)^2\right)^2 \sqrt{\det G} + i m B^N +
iB^R\right)  ,
\end{equation}
which describes a $D$-string with tension $((m+\chi)^2 + 1/g^2)^{1/2}$.

\vskip 0.3cm
\noindent{\it Other branes}
\vskip 0.2cm

The case $p=3$ is described in the absence of a condensate of the
open-string
potential by $M^3_{IJ}= (-I_4,  I_4)$  and
$M^3_{ab}=i(\gamma^1\gamma^2\gamma^3\gamma^4)_{ab}$, which is
self-dual in the
eight-dimensional transverse space.
In the presence of a non-zero condensate the  world-volume theory for the
three-brane is given by 4-dimensional $N=4$ supersymmetric
Yang-Mills theory.  The
theory is invariant under electric-magnetic duality transformations
when these
are accompanied by $SL(2,Z)$ transformations acting on the
closed-string bulk fields
 \cite{tseytlina,greengutperle}.

The cases with $p=5$ and $p=7$ are the magnetic duals of the
$D$-string and the
$D$-instanton, respectively.  The rotations that define the
boundary states
are obtained from the $p=1$ and $p=-1$ cases simply by multiplying
the matrix
$M_{ab}$ by $\gamma^1 \cdots \gamma^8$, which implements Hodge
duality in the
transverse eight-dimensional transverse space.

In the case of the branes of the  type IIA  theory the possible
values of $p$
are even and the left-moving and right-moving supercharges are
described by
$SO(8)$ spinors of opposite type.   Thus, the eight linearly realized
supersymmetries that annihilate the boundary state are
$(Q^a+M_{a\dot{a}}\tilde{Q}^{\dot{a}})$,
while the non-linearly realized ones are
$(Q^{\dot a}+M_{\dot{a} a}\tilde{Q}^{a})$.
The analysis is very similar to the Type IIB cases.

\section{Configurations with two or more  branes}

In  the sector with two separated $p$-branes the leading
contribution to the
free energy comes from the independent  energies defined by  a
functional
integral over  two disk world-sheets  -- one with Dirichlet conditions at
$y_1^i$ and the other at $y_2^i$.
The energy (per unit volume)  between these  separated branes is  determined
to lowest
order in
the string coupling constant by quantum fluctuations that describe
the exchange
of a closed string between them.  This is represented by a
world-sheet
diagram with the topology of a cylinder  with boundaries that
describe the
world-lines of the end-points of open strings moving in the
world-volume of
each brane.      The dependence of this energy on the  separation of the
branes, $L= |y_2 - y_1|$
determines the force between  them.  If the branes are viewed as
instantons the diagram represents the action per unit euclidean volume.

The general expression for the energy (or action) is given by
\begin{equation}
Z( F_1,F_2,L) =   C_1C_2 \int_0^\infty {dt\over 2p^+}
{_{(-1)}\langle} B,
y_2|  e^{-(P_{cl}^- - p^-) t}\hat R(M^{p_2 T}M^{p_1})
| B, y_1\rangle_{(-1)},
\label{genen}
\end{equation}
where $p^- = i\partial / \partial y^+$ and the value of $\eta$ is
taken to be the same for both end-states (both of them are either BPS
or anti-BPS).
With an appropriate choice of the rotations, $M^{p_1}$ and $M^{p_2}$, and
normalizations, $C_1$ and $C_2$ (\ref{genen})  represents the
energy between
branes of arbitrary $p_1$ and $p_2$, oriented in any relative
direction and with
arbitrary condensates, $F_1$ and $F_2$, of the  open-string  gauge
potentials
on
each brane (which are assumed to vary only slowly).  Furthermore, the
force between an anti--BPS state and
a BPS state is described  by rotating through $2\pi$  around all axes,
thereby  changing the sign of $\eta$ in
one of the states.   The normalizations $C_1(F_1)$   and $C_2(F_2)$
may be determined as before by considering the process to be a trace
over the open-string states that propagate around the annulus with
their  end-points fixed in the respective
world-volumes   of the branes. Writing $Z$ as a product of zero-mode
and non zero-mode factors,
\begin{equation} Z( F_1,F_2,L)
 \equiv  Z_{0}(F1,F2)Z_{osc}(M^{p_2T}(2) M^{p_1}(1)),
\label{zfact}
\end{equation}
it is clear that $Z_{0}(F1,F2,L)$ depends explicitly on $F_1$ and
$F_2$ separately while $Z_{osc}(M^{p_2T}(1)
M^{p_1}(1))$ depends only on the relative rotation.

\subsection{ Parallel branes}

The fact  that  D-branes  are BPS configurations means that two
parallel branes
of the same type  (the same values of $\eta$ and of the open-string
condensate
fields, $f_\alpha$ on
each brane) do not exert a force on each other.   The force  vanishes
identically as follows from the fact that exactly half
the space time supersymmetry is broken  by this configuration, which
leads to the vanishing of this amplitude and all higher contributions
with the topology of a sphere with $n$ holes cut out. This was
observed in the
case in which all the boundaries are Dirichlet in \cite{greeng}, in which
case the \lq energy' (actually, the action per unit volume) is
\begin{equation}\label{minusoinee}
Z(0,0,L) = \int_0^\infty {dt\over 2p^+}
{_{(-1)}\langle} B, y_2|  e^{-(P^-_{cl} -   p^-) t}
| B, y_1\rangle_{ (-1)},
\end{equation}
which vanishes because of the  cancellation between the exchange of
massless
states in the \NSNS\ and the \RR\  sectors,
\begin{equation}\label{exchange}
\langle \tilde I | \langle  I   |   J \rangle |\tilde J\rangle   -
\langle
 \tilde{\dot{a}}
| \langle \dot{a} |  \dot{b }\rangle |\tilde{ \dot{b}\rangle}=8-8=0.
 \end{equation}
  For widely separated $D$-instantons this  vanishing is attributed to a
cancellation between the  exchanged  massless scalar and
pseudoscalar states
\cite{greenc}.    This  generalizes to the case of parallel
$D$-branes of the same type with arbitrary $p$ and with the same boundary
condensates  \cite{polchina} simply by inserting $1 = R^\dagger(M^p)
R(M^p)$ in the
matrix element and using the expression (\ref{rotins}).  The
normalization
factor $C_1C_2$ is finite so that the energy vanishes in these
cases also. The
tension of the brane can be deduced by isolating  the \NSNS part of
(\ref{exchange}).

However, the interaction energy between two branes of opposite type (one being
BPS and the
other anti-BPS)  does not vanish. Since a BPS state is converted into an
anti-BPS state by a $2\pi$ rotation around all axes, $|B,-\rangle = \hat
R(M {(2\pi)})|B,+\rangle$,  the cylinder diagram that describes the
interaction  of a BPS $p$-brane  and an anti-BPS $p$-brane is given by,
\begin{eqnarray}\label{cylinder1}
Z_{+-}(0,0,L) &=&\int_0^\infty dt \langle B,  y_2, +|   e^{- 2p^+ (P^- -
p^-) t} \hat R(M {(2\pi)})|
B,y_1, +\rangle\nonumber\\
&=& \int_0^\infty dt
(\pi^2t)^{ (p-9)/ 2}e^{- (y_1-y_2)^2 /4\pi t}  {\prod_{n>0}(1+q^{2n})^8\over
\prod_{n>0}(1-q^{2n})^8}.
\end{eqnarray}
where $q=e^{-\pi t}$
(a special case of this with purely Dirichlet boundary conditions,
$p=-1$, was
considered  in \cite{greeng}).

The cylinder diagram is transformed to an annulus by a modular
transformation, in which case the expression for $Z$ is interpreted in
terms of a trace over the states of open strings with end-points fixed
in the world-volumes of the two branes.   This is the appropriate
coordinate frame for describing the   singularities of this process as
a function of the separation, $L=|y_2-y_1|$.    After the standard
change of variables,   $t' =1/t$ and $w=e^{-\pi t' }$,
(\ref{cylinder1}) becomes,
\begin{equation}
Z_{+-}(L) ={\pi^{ (p-9)} \over 16}\int {dt' \over t' }
{t'}^{- (p+1)/2}\exp\left(-t'  {1\over {4\pi}} (y_1-y_2)^2+\pi t' \right)
\frac{\prod(1-w^{2n-1})^8}{\prod(1-w^{2n})^8}.
\end{equation}
 The expression possesses
singularities  at
$L^2 =4 \pi^2 (1-n)$.  The $n=0$ term gives a pole at space-like
separation while the
$n=1$ term is a singularity at null separation and for $n>1$ the
singularities
are at time-like separations.   This is the $D$-brane
generalization of observations concerning the analytic structure of
the purely
Dirichlet amplitudes relevant to the $D$-instanton
(\cite{greeng}
and references  therein).   For the $p$-brane interpretation
$(y_2-y_1)$  is
necessarily space-like and only the $n=0$ term leads to  a physical
singularity
at  $L^2 = 4\pi^2$ \cite{susskinda}.  This represents the onset of a
tachyonic instability in the system.  In this situation it requires
more detailed dynamical understanding to determine
 whether there might be a bound state.

In the  light-cone frame
appropriate to the
$(p+1)$-instanton
interpretation $(y_2-y_1)^2$  has indefinite sign and the process has
singularities on and inside the light-cone as well as a single
singularity
outside the light-cone.

\subsection{Parallel $D$-strings with   boundary condensates}

When the condensates $F_1$ and   $F_2$ are different on two otherwise
identical parallel branes  supersymmetry is broken because the combination of
supercharges that annihilates one  boundary (\ref{susyi})  is
different from
the combination that annihilates the other.  This  leads to an
interaction  energy
between the branes.  The large-distance behaviour of
this energy is dominated by the $\tau\to  \infty$  limit which
picks out
the massless modes of the closed-string exchanged between the
branes.

For example, in the case of two parallel $D$-strings ($p=1$)  the
coefficients in
(\ref{genen}) are given by $C_1 = \sqrt{1+f_1^2}$ and  $C_2 =
\sqrt{1+f_2^2}$.  The zero mode factor in $Z$ is given by,
\begin{eqnarray}\label{angles}
Z_{0}(F_1,F_2,L)&=& \langle B_0, F_1, \eta | B_0, F_2,\eta
\rangle\nonumber \\
 &  =& C_1 C_2(
M_{IJ}^T(2)M_{JI}(1) -
M_{ab}(2)M_{ba}(1)) \nonumber \\
&=&\frac{8+4(f_1^2+f_2^2)+8f_1^2f_2^2+8f_1f_2}
{\sqrt{(1+f_1^2)(1+f_2^2)}}-8(f_1f_2+1).
\end{eqnarray}

In the case in which $f_1=f_2$ this
vanishes as the energy between two identical BPS-saturated branes should.
More generally, the expression (\ref{angles}) has the property that
for small $f_1$ and $ f_2$ it behaves as
\begin{equation}\label{smallf}
Z_0(f_1,f_2,L) \sim f_1^4 + f_2^4 + 6f_1^2 f_2^2  - 4f_1^3 f_2 -
4f_2^3 f_1 + O(f^5).
\end{equation}
The lower order terms  vanish which corresponds to the fact
that the moduli space of parallel $D$-strings is flat  \cite{bachasb}.

The oscillator factor, $Z_{osc}$,  in  (\ref{zfact})  can  be
evaluated exactly  using
(\ref{rdef}).  Since  the right-moving degrees of freedom do not
appear in the
rotation operator operator  the expression reduces to a trace over the
left-moving oscillator states.  The  generators
(\ref{gendefs}) can be
diagonalized by a unitary transformation,
\begin{eqnarray}\label{diag}
\alpha_n^I\to U_{IJ}\alpha_n^J\;&,&\;\alpha_{-n}^I\to U_{IJ}^\dagger
\alpha_{-n}^J \nonumber \\
S_n^a\to V_{ab}S_n^b\;&,&\;S_{-n}^{a}\to V_{ab}^\dagger S_{-n}^b.
   \end{eqnarray}
where $U$ and $V$ are defined  so that
\begin{eqnarray}\label{unitra}
U^\dagger T^{(\alpha)}    U&=&\pmatrix{i\alpha & 0&0\cr
     0 & -i\alpha & 0\cr
    0& 0& 0_6\cr}\nonumber \\
V^\dagger T^{(S)} V &=& \pmatrix{-i\half \alpha I_4 & 0 \cr
    0 & i\half \alpha I_4\cr},
\end{eqnarray}
where $\alpha$  (defined in (\ref{cosalph})) is the angle of the
relative rotation $M_1^TM_2$.
The traces over the oscillator states are standard  and the result
is given by
\begin{equation}\label{part}
Z_{osc}(\alpha,L)  = \pi^4
\int_0^\infty dt t^{-4}
\frac{\prod_{n=1}^\infty (1-q^{2n}e^{i\alpha/2})^4 (1-q^{2n}
e^{-i\alpha/2})^4} {\prod_{n=1}^\infty (1-q^{2n})^6(1-q^{2n}e^{i\alpha})
(1-q^{2n}e^{-i\alpha})}\exp\left(-{L^2 \over {4\pi t}}\right),
\end{equation}
where   $q=e^{-\pi t}$ and  the zero-mode factor follows from
(\ref{angles}).
This can be written in a compact
form in terms of Jacobi $\theta$ functions and the Dedekind $\eta$
function,
\begin{equation}
Z_{osc}(\alpha,L) ={\pi^4 \over 16} {\sin(\alpha/2)\over
  \sin^4(\alpha/4)} \int dt t^{-4}\frac{\theta_1^4
(\frac{\alpha}{4\pi}\mid
it)}{\eta^9(it)\theta_1(\frac{\alpha}{2\pi}\mid it )}\exp\left(-{L^2 \over
{4\pi t}}\right),
\end{equation}
This expression exhibits  interesting analytical structure  as
a function
of the separation  which is exposed in the short cylinder limit,
 $t\to
0$.  The Jacobi transformation,  $t \to t^\prime=1/t $, transforms
the cylinder  into an annulus, giving,
\begin{equation}\label{modpart}
Z_{osc} (\alpha,L) ={\pi^4 \over 16} {\sin(\alpha/2)\over
  \sin^4(\alpha/4)} \int
\frac{dt^{\prime}}{t^\prime}
\frac{\theta_1^4(-it^\prime\frac{\alpha}{4\pi}\mid
it^\prime)}{\eta^9(it^\prime)\theta_1(-it^\prime\frac{\alpha}{2\pi}\mid
it^\prime)}\exp(-t^\prime{L^2 \over {4\pi }}).
\end{equation}

The complete expression for $Z$  (\ref{zfact}) is given by the
product of (\ref{angles}) and (\ref{modpart}).
 The argument  of the $\theta_1$ is imaginary so
that in the short cylinder
limit, $t^\prime\to \infty$,  the  behaviour of the integrand of
(\ref{modpart})  is  given by
\begin{eqnarray}
\lim_{t^\prime\to\infty}\frac{ e^{-t'L^2/4\pi}
\theta_1^4(-it^\prime\frac{\alpha}{4\pi}\mid
it^\prime)}{\eta^9(it^\prime)\theta_1(-it^\prime\frac{\alpha}{2\pi}\mid
it^\prime)}&\sim &\lim_{t^\prime\to\infty}\frac{e^{-t'L^2/4\pi} \sin^4(-i\alpha
t^\prime/4)}{\sin(-i\alpha/2
t^\prime)}\nonumber \\
& \sim &\lim_{t^\prime\to\infty}e^{-t'(L^2 - 2\pi |\alpha| )/4\pi} + O(e^{-\pi
t^\prime}).
\end{eqnarray}

Hence,  the expression
diverges
for separations
$L^2 = (y_2-y_1)^2 \le 2\pi\mid\alpha\mid$. If the transverse space has
Minkowski
signature  this
indicates the presence of  a pole in $L^2$ outside the light-cone
as well as an
infinite set of poles at values
\begin{equation}
L^2=2\pi(\mid\alpha\mid-2\pi n),\quad n=0,1,\cdots
\end{equation}
Note that (\ref{cosalph}) implies that the limit of infinite field
strength corresponds to
$\alpha=\pi$.

After continuing to Minkowski signature the magnetic boundary
condensate turns into an electric boundary condensate and the rotation
matrices become  boosts. This change of signature can be accomplished
by $f\to i f$ which implies $\alpha \to i\alpha$. The argument of the
$\theta$ functions are then real and the integrand of
(\ref{modpart}) exhibits an
infinite number of simple poles on the real $t'$ axis at
$t'=2\pi(2k+1)/\alpha$ where $k $ is integer.  This leads to an
imaginary part in the energy, which is given by the sum over the
residues of the poles.  This is very similar to the situation  in
\cite{Bachas},  where the type I superstring was considered, which has
Neumann boundary conditions in all directions.  In fact, the rather
complicated sum over spin structures obtained in \cite{Bachas} reduces
to a simple expression of the form  (\ref{modpart}).

\subsection{Parallel, intersecting and orthogonal   branes.}

 The most  general configuration  of two branes is one in which the
branes may have different values of $p$, they may
both be BPS states or one of them may be anti-BPS and they may
or may not be parallel.  We will be interested in describing which
configurations have residual supersymmetries.   The
important issue  of determining whether there are BPS bound states of
branes lying within one another will not be addressed in the
following.  Furthermore, we will not consider
the rich class of configurations in  which there is a non-trivial
boundary condensate.

A systematic way of analyzing configurations is to
first consider
one of the branes to be a $D$-instanton  ($p_1=-1$) and the other
one to have
$p_2 = -1,1,3,5$ or $7$ (the type IIA case is analogous but $p_1$
and $p_2$
take even values).  All other cases can then be obtained  by  appropriate
rotations.

\subsubsection{ A $D$-instanton and a $p$-brane.}

The zero-mode contribution to the cylinder diagram with $p_1 =-1$  with general
$p_2$ is
\begin{eqnarray}\label{rotonetwo}
Z_0(p_1 = -1, p_2) &= &{_{(p_2)}\langle} B_0 | B_0\rangle_{(-1)} =
{_{(-1)}\langle}
B_0 | R_0^{p_2T} |B_0\rangle_{(-1)} \nonumber\\
 &= &  \Tr M^{p_2}_{IJ} - \Tr M^{p_2}_{ab}.
\end{eqnarray}
The unbroken supersymmetry generators that annihilate the two
boundary states
are given by
\begin{equation}\label{twoqs}
Q_{(-1)}^{+a} = Q^a + i \tilde Q^a, \qquad
Q_{(p_2)}^{+a} = Q^a + i M_{ab}^{p_2} \tilde Q^b,
\end{equation}
(with similar combinations of the dotted supercharges) where the
cases of $\eta
=\pm 1$ are included by allowing  for the fact that the overall sign of
$M_{ab}^{p_2}$ can be changed by a $2\pi$ rotation in all
directions.   The
compatibility of these two supersymmetries depends on the value of
$p_2$.  The
cases $p_2$ and $p'_2 = 6-p_2$ have similar behaviour.

Thus, $Z_0 =0$ when $p_2 = -1$ or $p_2 = 7$ (the rotation matrix
$\gamma^1 \cdots \gamma^8 \equiv 1$) due to the
cancellation between
the \NSNS\ and \RR\ sector as described earlier.   In this case both
supercharges are identical and there is a total of 16 unbroken
supersymmetries.

For $p_2 =3$ the two traces in (\ref{rotonetwo}) vanish {\it
individually} so
that
once again $Z_0=0$, but this time due to the separate vanishing of
the  \NSNS\
and \RR\ sectors.  The two matrices $M^{p_r}_{ab} (r)$ (where $r=1,2$ labels
the two branes) are of the form
\begin{equation}
M^{-1}_{ab}(1) =\delta_{ab}, \qquad M^{3}_{ab}(2) =\gamma^{1234}_{ab}.
\end{equation}
Since $(\gamma^{1234})^2=1$, $M^3_{ab}$ has eigenvalues $= +1,-1$. The
projector of brane 2  on the eigenvectors with eigenvalue +1 is given by
\begin{equation}
P_{ab} (2) =\frac{1}{2}(1-\gamma^{1234})_{ab}.
\end{equation}
The dimensionality of this eigenspace is given by the trace of the
projector $\tr P(2) =4$. Therefore, there are four common eigenvalues of
$M^{(-1)}_{ab}(1)$ and $M^3_{ab}(2)$ and so  four undotted
supersymmetries   remain
unbroken. The same reasoning also applies to the dotted
supersymmetries.  Hence the total number of unbroken supersymmetries
is eight  -- the configuration preserves a quarter of the original
supersymmetry.   Furthermore, this configuration is    invariant
under rotation by $2\pi$ that reverses the overall sign of $M^{3}_{ab} $,
which interchanges the BPS and anti-BPS conditions.

When $p_2=1$ or $p_2 = 5$  the bi-spinor trace vanishes but the
bi-vector trace
is non-zero so that $Z_0 \ne 0$.   In this case $M^{1}_{ab} =
\left(\gamma^1\gamma^2\right)_{ab} $ has imaginary eigenvalues so that
the conserved supercharges of the two branes have no eigenvalues in
common  -- there are no unbroken
space-time supersymmetries in this case.  This means that there is a
force between the branes so that they might form a bound state that can
be a BPS state \cite{polchinskinotes,douglasa}.

 This behaviour of the zero modes extends to the non-zero mode factor,
$Z_{osc}$.
The non-zero
mode  contributions  from fermionic and bosonic oscillators cancel
in the cases
$p_2 = -1,3,7$, so that  $Z_{osc} = 1$.  In
the cases
$p_2 = 1,5$  $Z_{osc}$   is a
non-trivial
ratio of contributions from bosonic and fermionic oscillators.

\subsubsection{ Parallel, intersecting and
orthogonal   $p_1$
and $p_2$ branes.}

The cases with $p_1 > -1$ can be obtained from the above by
inserting $1 =
R^{(p_1) T} R^{(p_1)}$ in (\ref{rotonetwo}), where $R^{(p_1)}$ is a rotation
that acts on the
bi-vector and bi-spinor indices so that,
\begin{equation}\label{rotm}
M_{IJ}^{p_1} R_{JK}^{(p_1)} = M_{IK}^{-1}, \qquad  M_{ab}^{p_1}
R_{bc}^{(p_1)} =  M_{ac}^{-1}.
\end{equation}
This is a $T$-duality transformation that changes the number of
directions in
which there are Neumann and Dirichlet boundary conditions.   In
this way the
general case can be reduced to a consideration of the ones
considered in (a)
above with $M^{p_2}$ replaced by $M^{\prime p_2} = R^{(p_1)} M^{p_2}$.

The relative orientation of the two branes depends on the choice of
$R^{(p_1)}$.  There are several distinct classes to consider.
\begin{itemize}
\item{(a)}   { \it Parallel branes}.  The $(p_2+1)$ world-volume
coordinates
  are a subset of the $(p_1+1)$ world-volume coordinates (where $p_2<p_1$).
\item{(b)} {\it Intersecting branes}.  A subset of the coordinates
of one brane
are orthogonal to a subset of the world-volume coordinates of the other.
\item{(c)}  {\it Orthogonal branes}.  The world-volume coordinates
of one brane
are all orthogonal to those of the other.  This is not possible for
the usual
$p$-branes since they always share the time direction -- but, it is
possible
for the $(p+1)$-instantons  considered in this paper, in which the
branes have
euclidean world-volumes.
\end{itemize}

As a specific example, consider the case of a $D$-string ($p_2=1$) and a
three-brane ($p_1 =3$).  Choosing the three-brane world-volume to
be in the
directions $I=1,2,3,4$, the three distinct configurations of the
$D$-string  correspond to the
choices for
the $p_2$ matrices,
\begin{eqnarray}
M_{IJ}^{p_2}  = {\rm diag} (-1,-1, 1, 1, 1, 1, 1, 1), \qquad M_{ab}^{p_2} =
\gamma_{ab}^{12},  \label{parallel}\\
M_{IJ}^{p_2}  = {\rm diag} (-1, 1, 1, 1,-1, 1, 1, 1), \qquad M_{ab}^{p_2} =
\gamma_{ab}^{15},  \label{inter} \\
M_{IJ}^{p_2}  = {\rm diag} ( 1, 1, 1, 1,-1,-1,  1,  1), \qquad
M_{ab}^{p_2} =
\gamma_{ab}^{56},
\label{orthog} \end{eqnarray}
which describe the parallel, intersecting and orthogonal cases,
respectively.

These may now be transformed into the case in which $p_1 =-1$ by
inserting the rotation $R^{(p_1)}$ that transforms the three-brane
into the
$D$-instanton,
\begin{equation}\label{rotthr}
R_{IJ}^{(p_1)} = {\rm diag} (-1,-1,-1,-1,1,1,1,1), \qquad
R^{(p_1)}_{ab} =
\gamma_{ab}^{1234}.
\end{equation}
Acting on the parallel $D$-string (\ref{parallel})  the result is another
$D$-string (the matrices $M^{p_2}$ transform into those of the
$p=1$ case).
Acting on the orthogonal $D$-string (\ref{orthog}) the result is a
$p_2 = 5$
state.  Both of these cases therefore break all the space-time
supersymmetries.
 However, the result of acting with $R^{(p_1)}$  on the intersecting $D$-string
(\ref{inter}) is a
$p_2 =3$ state and so one quarter of the supersymmetries (i.e., 8) are
preserved.  This
is in accord with the results in \cite{stromingerb,polchinskinotes}.

Another illustrative example is the case with $p_1=1$ and $p_2=5$
where there are three  possible distinct configurations of the string and the
fivebrane.
  In the
intersecting case, in which there is one common world-volume
direction, no
supersymmetries are conserved.  The parallel configuration in which
there are
two common world-volume directions preserves one quarter (i.e., 8) of the
supersymmetries.  The case in which the world-volumes are
orthogonal preserves
half (i.e., 16) the  space-time supersymmetries, note that the last
configuration has no analogue for the p-branes which share the
world-volume time direction.

The general result coincides with that of Polchinski's   argument
 \cite{polchinskinotes}.  There, it was shown
that the number  of
directions in which the boundary conditions at either end of the
string are
different must be $0$ mod $4$ if any space-time supersymmery is
preserved.
This coincides with the number of $-1$ entries in the diagonal matrix
$M_{IJ}^{ p_1} M_{JK}^{p_2}$.

The following table summarizes the number of surviving  dotted and undotted
supersymmetries for all
$p_1$ and $p_2$  relevant to the type IIB theory.
The integer N denotes the number of common Neumann
directions of the (euclidean) $(p_1+1)$-dimensional and $(p_2+1)$-dimensional
world-volumes.  The
values marked  $n/a$ require a total of more than eight dimensions
transverse to $x^\pm$ and cannot be described in our light-cone frame.  The
values marked - do not exist.

\vskip 0.5cm

{\bf Type IIB}\hfill\break\indent
\begin{tabular}{|c||c ||c |c|c||c|c|c|c|c||c|c|c||c|} \hline
$p_1$&\multicolumn{1}{c||}{ -1}&\multicolumn{3}{c||}{
  1}&\multicolumn{5}{c||}{ 3}&\multicolumn{3}{c||}{
5}&\multicolumn{1}{c|}{
7}\\ \hline

 N&0& 2&1&0&4&3&2&1&0&6&5&4&8 \\ \hline\cline{1-14}
$p_2=-1$&16&- &- &0 &- &- &- &- &8 &- &- &- &- \\ \hline
$p_2=1$&0&16 &0 &8 &- &- &0 &8 &0 &- &- &- &- \\ \hline
$p_2=3$&8 & 0&8 &0 &16 &0 &8 &0 &16 &- &- &- &- \\ \hline
$p_2=5$&0&8 &0 &16 &0 &8 &0 &n/a &n/a &16 &0 &8 &- \\ \hline
$p_2=7$&16&0 &n/a&n/a &8 &n/a&n/a &-&-&0 &n/a &n/a &16 \\ \hline
\end{tabular}
\vspace{1.0cm}

The type IIA theories can be described similarly and the following table
summarizes the surviving supersymmetries in that case:
\vskip 0.25cm
{\bf Type IIA}\hfill\break\indent
\begin{tabular}{|c||c  |c||c|c|c|c||c|c|c|c||c|c|} \hline
 $p_1$ &\multicolumn{2}{c||}{ 0}&\multicolumn{4}{c||}{
2}&\multicolumn{4}{c||}{
4}&
 \multicolumn{2}{c|}{ 6}\\ \hline

 N& 1 &0&3&2&1&0&5&4&3&2&7&6\\  \hline\cline{1-13}
$p_2=0$& 16  &0&- &- &0 &8 &- &-     &-     &-     &- &- \\ \hline
$p_2=2$& 0 &8 &  16&0    &8    &0&- & -    & 0    & 8    &- &- \\ \hline
$p_2=4$& 8   &0 &0 &8     &0    &16 &16 &0    &8    &0   &- &-  \\ \hline
$p_2=6$& 0   &16 &8     &0    &n/a&n/a &0    &8    &n/a   & n/a &16 &0 \\
\hline
\end{tabular}
\vspace{2.0cm}

\subsubsection{ Configurations with more than two branes.}

When there are three  or more $D$-branes a smaller fraction of the
supersymmetries can be preserved by the solutions.

\vskip0.3cm
(a) {\it 4 residual supersymmetries}
\vskip 0.2cm

A simple example is provided by the configuration of three orthogonal
$D$-strings with world-sheet  orientations in the $(1,2);(3,4);(5,6)$
directions. The boundary conditions are defined by the matrices,
\begin{eqnarray}
M^1_{IJ}(1)& =& {\rm diag} (-1,-1,1,1,1,1,1,1), \qquad M^1_{ab}(1)
=\gamma^{12}_{ab}\\
M^1_{IJ}(2)& =&{\rm diag} (1,1,-1,-1,1,1,1,1), \qquad M^1_{ab}(2)
=\gamma^{34}_{ab}\\
M^1_{IJ}(3)& =& {\rm diag}(1,1,1,1,-1,-1,1,1), \qquad M^1_{ab} (3)
=\gamma^{56}_{ab}.
\end{eqnarray}
The \NSNS matrices may be transformed by T-duality into those of a
$D$-instanton  and two three-branes with world-volumes that share two
directions,  $M^{-1}_{IJ}(1)$, $M^3_{IJ}(2)$
and $M^3_{IJ}(3)$.  The \NSNS matrices may be transformed by T-duality
into those of a
$D$-instanton  and two three-branes with world-volumes that share two
directions,  $M^{-1}_{IJ}(1)$, $M^3_{IJ}(2)$
and $M^3_{IJ}(3)$.  The
\RR\ matrices transform under this  T-duality to
\begin{equation}
M^{-1}_{ab}(1) = \delta_{ab}, \qquad M^{3}_{ab}(2)=\gamma^{1234}_{ab},
\qquad M^{3}_{ab}(3)=\gamma^{1256}_{ab} .
\end{equation}
The common supersymmeties are the ones which have eigenvalues
$+1$. There are two projectors onto the eigenspaces with eigenvalue
$+1$ for $M_{ab}^3(2)$ and $M^3_{ab} (3)$, respectively,
\begin{equation}
P_{ab} (2) =\frac{1}{2}(1-\gamma^{1234})_{ab}, \qquad P_{ab} (3) =
\frac{1}{2}(1-\gamma^{1256})_{ab}.
\end{equation}
Since $[P(2),P(3)]=0$ the product
$P(2)P(3)$ is
also a projector
given by
\begin{equation}
\left(P(2)P(3)\right)_{ab}=\frac{1}{4}(1-\gamma^{1234}-\gamma^{1256}
 - \gamma^{3456})_{ab}.
\end{equation}
Since $\Tr(P(2)P(3))=2$ there are two unbroken undotted
supersymmetries.  Together with  the dotted supersymmetries there is a
total of  four unbroken  supersymmetries -- this configuration
preserves $1/8$  of the
original  $32$ supersymmetries.

If this configuration of three $2$-instantons is compactified on
$T^6$ in the $(1,2,3,4,5,6)$ directions the result is an
instanton in four dimensions which leads to the
non-conservation of the three  axionic scalar \RR charges associated with
$B^R_{12}$,$B^R_{34}$,$B^R_{56}$.

Another example is a  four-dimensional black hole state that can be
made by combining  three
self-dual three-brane world-volumes in the orientations
$(1,2,3,4);(1,2,5,6);(1,3,5,7)$, interpreting the $1$ direction as
the time axis and compactifying the directions $2-7$ on $T^6$.  Since
the three-brane does not couple to the dilaton this configuration must
correspond to a conventional  Reissner--Nordstrom black hole of the
type IIB theory.  A fourth three-brane in the $(1,4,6,7)$ direction
can  also be added.  Any of these branes shares its three spatial axes
with one of the other ones -- this is a stable configuration.

Analogous statements also apply for the Type IIA theory in an
obvious manner \cite{townsendpapa}.

\vskip 0.3cm
{\it 2 residual supersymmetries}
\vskip 0.2cm

There are also configurations which only preserve 1/16 of the original
supersymmetry. One example in  the type IIB case is a
three-brane with world-volume oriented in the  $(1,2,3,4)$ directions
and three mutually orthogonal $D$-string world-sheets in the
$(1,5);(2,6);(3;7)$
directions  which each intersect the
three-brane in one  direction. After performing a
T-duality transformation that converts the three-brane to a
$D$-instanton,   the projectors for the three one-brane supersymmetries
which will  be compatible with the unbroken
supersymmetries of  the three-brane are given by,
\begin{eqnarray}
P(2)&=&{1\over 2}(1-\gamma^{2345})\\
P(3)&=&{1\over 2}(1+\gamma^{1346})\\
P(4)&=&{1\over 2}(1-\gamma^{1247}).
\end{eqnarray}
These three projectors commute and their  product is
also a projector, $P$, satisfying
\begin{equation}
\tr P= \tr(P(2)P(3)P(4)) = 1,
\end{equation}
so that there is one undotted unbroken
supersymmetry, as well as a similar dotted one.  This gives a total of
two unbroken supersymmeties.  Other similar configurations can easily
be constructed.

\vskip 0.3cm

\section{Linear and non-linear realizations of space-time supersymmetry}

Since a Dirichlet $p$-brane breaks translational
invariance in the
directions transverse to the brane as well as
half the space-time supersymmetries \cite{hughespolchinskib} there
must be a  supermultiplet of eight bosonic zero modes (which form a
world-volume $p$-component vector) and eight fermionic zero modes
living in its  world-volume.
In the case of brane-like solitonic solutions of
type IIB  supergravity theories these are modes of the bulk
fields \cite{callanstromb,duffluc} while in
the $D$-brane description they are the ground states
of the open string.

The unbroken symmetries are realized linearly on the fields
whereas the broken symmetries are realized nonlinearly.   There is the
possibility for some confusion here  since in  the light-cone frame
half the 32
components of the supersymmetry in the type IIB theory are already
realized
non-linearly.   They  divide into two groups of 16.  One of these
groups, $Q^a$
and $\tilde Q^{  a}$ (associated with the $SO(8)$ spinor parameters
$\epsilon^a$ and $\tilde \epsilon^a$),  is realized linearly while
the other 16
components, $Q^{\dot a}$ and $\tilde Q^{\dot a}$ (associated with
the $SO(8)$
spinor parameters $\eta^{\dot a}$ and $\tilde \eta^{\dot a}$), is
realized
non-linearly.

Now we wish to  divide these 32 components into groups of 16 in a
different
manner that accounts for the fact that the boundary state links
left-movers and
right-movers.
The boundary state is annihilated by the  combination of
supercharges  given in  (\ref{susyi}),
\begin{equation}
Q^{+\;a}\mid B\rangle=Q^{+\;\dot{a}}\mid B\rangle=0,
\end{equation}
which are  the unbroken supersymmetries while $Q^{- b},Q^{-
\dot{b}}$ define
the broken supersymmetries.  The parameters of the corresponding
transformations will be denoted  $\eta_\pm^a = \eta^a \pm \tilde
\eta^a$ and
$\epsilon^{\dot a}_\pm = \epsilon^{\dot a} \pm \tilde
\epsilon^{\dot a}$.

The algebra of the broken and unbroken supercharges is given by the
following
commutation
relations
\begin{eqnarray}\label{plusminc}
\{Q^{+ a},Q^{- b}\} = 2\delta^{ab}p^+&& \qquad  \{Q^{+ a},Q^{-
\dot{b}}\} =
{\gamma^I_{a\dot{b}} \over \sqrt 2} (p^I+M^{IJ}\tilde{p}^J)\nonumber \\
\{Q^{- \dot{a}},Q^{- \dot{b}}\} &= &
\delta^{\dot{a}\dot{b}}(P^-+\tilde{P}^-) = \delta^{\dot a\dot b} P^-_{cl},
\end{eqnarray}
with all other anticommutators vanishing.

The open-string  light-cone vertex operators  are defined only in
the case that
the momentum carried by the emitted state satisfies $k^+=0$ (our
conventions
will follow those of \cite{greenseiberg}). In this case the physical state
condition $\zeta \cdot k$ (where $\zeta^\mu$ is the  physical open-string
vector potential) can be satisfied by choosing  special transverse
polarizations satisfying $\zeta^I k^I =0$, and  $\zeta^- = \zeta^I
k^I /k^+$
can
be arbitrary.   Similarly,  the physical state condition on the
ground-state
open-string spinor field (which has $SO(8)$ components $u^a$,
$u^{\dot a}$) is
$u^a = -2k^I \gamma^I_{a\dot a} u^{\dot a}/k^+$ which can be
satisfied  with
finite $u^a$ by choosing the  $ \gamma^I_{a\dot a}k^I u^{\dot a} =0 $.

In order to avoid awkward singular expressions in the supersymmetry
transformations  it is convenient to take $u^{\dot a}$ to be
proportional to
$k^+$ and define $v^{\dot a} = u^{\dot a}/k^+$, which remains
finite as $k^+\to
0$.   The relation between the two $SO(8)$ components of the ground-state
fermion becomes $u^a=-k^I\gamma^I_{a\dot{a}}v^{\dot{a}}$.    The
independent
wave functions are then taken to be $\zeta^I$ and $v^{\dot a}$.

Supersymmetry transformations on the $SO(8)$ components of the massless
open-string fields take the form,
\begin{eqnarray}\label{susytr}
\tilde{\zeta}^I&=&  \eta^a\gamma^I_{a\dot{a}}u^{\dot{a}} \sim 0 ,\qquad
\tilde{\tilde{\zeta}}^I = \sqrt{\frac{1}{2}} \epsilon^{\dot{a}}
\gamma^I_{a\dot{a}}u^a + \frac{\sqrt{2}} {k^+}
\epsilon^{\dot{a}}u^{\dot{a}}k^I
=
k^I \epsilon^{\dot a} \gamma^{IJ}_{\dot a \dot b} v^{\dot b} ,
\nonumber \\
 \tilde{v}^{\dot{a}} &=&   \eta^a\gamma^I_{a\dot{a}}\zeta^I , \qquad
\tilde{\tilde{v}}^{\dot{a}}={1\over k^+} \sqrt{\frac{1}{2}} \left(
\epsilon^{\dot{a}}\gamma^{IJ}_{\dot{a} \dot{b}}k^I\zeta^J+
\epsilon^{\dot{a}}
\zeta^Ik^I\right)
\end{eqnarray}
(the apparent singularity in the last term does not contribute to the
expressions for the on-shell vertex operators).

The massless  bosonic open-string vertex operator  is
given by
\begin{equation}\label{bosvert}
V_B(\zeta,k) = (\zeta^IB^I-\zeta^-)e^{ikX},
\end{equation}
where
\begin{equation}\label{bosfact}
B^{I}  = \partial X^I-\frac{1}{2}S^{a}(z)\gamma^{IJ}_{ab}S^b(z)k^J.
\end{equation}
The massless fermion vertex operator is given
\begin{equation}\label{fermvert}
V_F(u,k) =  (u^aF^a+u^{\dot{a}}F^{\dot{a}})e^{ikX},
\end{equation}
where,
\begin{equation}\label{fermfac}
F^a = S^a(z) ,\qquad
F^{\dot{a}} =  \gamma^I_{a\dot{a}}S^{a}(z)\partial
X^I+\frac{1}{6}:
\gamma^I_{a\dot{a}}S^a(z)S^b(z)\gamma^{IJ}_{bc}S^c(z):k^J.
\end{equation}

The   32 components of the supersymmetry act on the
vertex operators in the following way,
\begin{eqnarray}\label{transvert}
\delta_\eta V_B=  [\eta^aQ^a,V_B(\zeta)]&=&V_F(\tilde{u}) \nonumber \\
\delta_\eta V_F= [\eta^aQ^a,V_F(u)]&=&V_B(\tilde{\zeta}) \nonumber \\
\delta_\epsilon V_B=
[\epsilon^{\dot{a}}Q^{\dot{a}},V_B(\zeta)]&=&V_F(\tilde{\tilde{u}})+
\epsilon^{\dot{a}}\partial_z W^{\dot{a}}_B(\zeta,k,z) \nonumber\\
\delta_\epsilon V_F=
[\epsilon^{\dot{a}}Q^{\dot{a}},V_F(u)]&=&V_B(\tilde{\tilde{\zeta}})+
\epsilon^{\dot{a}}\partial_z W^{\dot{a}}_F(u,k,z)
\end{eqnarray}

The total derivatives $W^{\dot{a}}_F,W^{\dot{a}}_B$ can
give rise to contact terms, which affect the discussion of
supersymmetry of the
bulk fields but will not enter to the discussion of the open-string
sector in
this paper. They are given by \cite{greenseiberg},
\begin{eqnarray}
W_B^{\dot{a}}&=&\sqrt{2}\gamma^{I}_{a\dot{a}}\zeta^IS^ae^{ikX}\nonumber\\
W_F^{\dot{a}}&=&\sqrt{2}v^{\dot{a}}e^{ikX} + \frac{\sqrt{2}} {8}
(\gamma^{IJ}u)^{\dot{a}} S^b \gamma^{IJ}_{bc} S^ce^{ikX}\label{totaldB}.
\end{eqnarray}

The open-string amplitudes can now be calculated in the cylinder
frame where
the width of the cylinder in $\sigma$  is $\pi$.  The cylinder
is chosen to
be semi-infinite with the boundary at $\tau=-\infty$ representing a
physical
closed-string state $\langle \Phi|$ that carries the non-zero
momentum $p^+$.
The open-string vertex operators are attached to the boundary at
$\tau=0$, and
the boundary conditions may be used to express them entirely in terms of
left-moving  operators.  These chiral vertices, which are integrated in
$\sigma$ along  are displaced infinitesimally away from the boundary at
$\tau=0$  with an arbitrary time ordering.   All permutations of
this ordering
are then summed with equal weight \cite{greengutperleb}.    The
amplitude with $n$
bosonic open-string ground states has the form,
\begin{equation}\label{openamplitude}
A_n(\Psi\mid \zeta_1,\cdots,\zeta_n)=\int
d\sigma_1..d\sigma_n\langle
\Psi\mid
V_B(\zeta_1,k_1,z_1)\cdots
V_B(\zeta_n,k_n,z_n)\mid B\rangle.
\end{equation}

A supersymmetry transformation of  this amplitude  is obtained  by
substituting
a transformed wavefunction $\tilde{\zeta_1}$ or
$\tilde{\tilde{\zeta_1}} $into
$A_n$. Using (\ref{transvert}) the transformed vertex operator can
be written
as a
commutator of a supersymmetry generator and a fermionic vertex
operator.  Thus, for  the linearly realized components of the conserved
supersymmetry the vertex,\begin{equation}\label{commutat}
V_B(\tilde{\zeta})=\eta^a_+ Q^{+ a}V_F(u)-V_F(u)\eta^a_+  Q^{+ a},
\end{equation}
can be inserted into (\ref{openamplitude}) and  the $Q^+$ in the
first term
acts to the left  on the closed string state $\Psi$ giving a
transformed state,
$\delta_{\eta_{\pm}}\Psi$.   The $Q^+$ in the second term is moved to the
right and gives transformed open string vertex operators until it hits
the boundary where it is annihilated  by the boundary state.   Thus the
conserved
 supersymmetry relates S-matrix elements with  $n+1$ bosonic states
(including
the one (bosonic) closed-string end-state)  to elements with $n-1$
bosonic and
2 fermionic states,
\begin{eqnarray}\label{nongold}
& A_n(\Psi\mid\tilde{\zeta}_1,\zeta_2,\cdots,\zeta_n)
=A_n(\delta_{\eta^+}
\Psi\mid u_1,\zeta_2,\cdots,\zeta_n) + A_n(\Psi\mid
u_1,\tilde{u}_2,\zeta_3\cdots,\zeta_n)\nonumber\\
&+ \cdots\nonumber +  A_n(\Psi\mid
u_1,\zeta_2,\cdots,\zeta_{n-1},\tilde{u}_n).
\end{eqnarray}
 This corresponds to the linearly realized
supersymmetry which is not broken by the boundary state.   A
similar analysis
applies to the non-linearly realized conserved supercharge,
$Q^{+\dot a}$ with
$\tilde \zeta$ and $\tilde u$ replaced by $\tilde{\tilde \zeta}$ and
$\tilde{\tilde u}$.

The supercharge $Q^{-a }$ is not annihilated by the boundary so
that  similar
manipulations for these supercharges leave a residual term.  This  is
proportional to  $\eta^a_- Q^{-a}|B\rangle$, which has the form of the
fermion emission vertex   (\ref{fermvert}) acting on the boundary, in which
the
supersymmetry parameter $\eta_-^a$ is the  wave function.
Therefore,  the
amplitude with $n+1$ bosonic states is related by the $Q^-$
supersymmetry  to a
sum of terms with $n-1$ bosons and two fermions, together with an
extra term
which has  an extra zero-momentum fermion insertion -- it has a
total of $n$
bosons and two fermions,
\begin{eqnarray}\label{goldstinoo}
& A_n(\Psi |  {\tilde{\zeta}}_1,\zeta_2,\cdots,\zeta_n)=
A_n(\delta_{\eta^-}
\Psi | u_1,\zeta_2,\cdots,\zeta_n) + A_n(\Psi  | u_1,
{\tilde{u}}_2,\zeta_3\cdots,\zeta_n)\nonumber\\
&+\cdots +  A_n(\Psi |  u_1,\zeta_2,\cdots,\zeta_{n-1}, {\tilde{u}}_n) +
A_{n+1}(\Psi\mid u_1,\zeta_2,\cdots,\zeta_n,\eta^{ - })   \end{eqnarray}
This  is the S-matrix statement of the nonlinear realization of  the
spontaneously broken $Q^-$ supersymmetry \cite{greengutperleb}.

 The
corresponding
analysis with the non-linearly realized supercharge $Q^{- \dot a}$
leads to the
same relationship between amplitudes but with $\tilde \zeta$ and
$\tilde u$
replaced with $\tilde{\tilde \zeta}$ and $\tilde {\tilde u}$.
Higher-order terms give rise to S-matrix elements with arbitrary
numbers of soft fermions.

Note that (\ref{mindef}) and (\ref{newthet}) implies
$Q^{-a}=\hat{\theta}^a$, hence the nonlinearly realized supersymmetry
charge $Q^{-a}$ acts as the modified Grassmann parameter
$\hat{\theta}^a$ on the boundary state $\mid B,+\rangle$ which is the
bottom component of a superfield expanded in $\hat{\theta}^a$.
The vertex
operator for the emission of zero momentum Goldstinos  is given by
$\eta^{-\;a}\theta^a\mid
B_0,\theta,+\rangle $, which indicates
that the nonlinearly realized supersymmetry acts as a shift on the
Goldstino field.

 One important subtlety in this analysis is that  contact terms
have to be
carefully accounted for.  These arise from the derivative terms in
the algebra
(\ref{transvert}) but in the open-string amplitudes considered here they
integrate to zero.  However, they are an important feature of more
general
amplitudes with  closed-string vertex operators.

\vskip 0.3cm
{\it Acknowledgments}
\vskip 0.2cm

MBG is grateful to the Physics Department at Rutgers University for
their hospitality while this work was completed and MG gratefully
acknowledges support by EPSRC and a Pannett Research Studentship of
Churchill College, Cambridge.

\end{document}